\documentclass[a4paper,fleqn,usenatbib,useAMS,letters]{mnras}

\usepackage{graphicx}	
\usepackage{amsmath}	
\usepackage{amssymb}	
\usepackage{multicol}        
\usepackage{bm}		

\newcommand{\rcw}{1E~161348$-$5055}
\newcommand{\rcwone}{RCW~103}
\newcommand{\Mdot}{\dot{M}}
\newcommand{\Omegas}{\Omega}
\newcommand{\Omegaej}{\Omegas_{\rm ej}}
\newcommand{\Pej}{P_{\rm ej}}
\newcommand{\tej}{t_{\rm ej}}
\newcommand{\tprop}{t_{\rm prop}}
\newcommand{\rmag}{r_{\rm m}}
\newcommand{\ralf}{r_{\rm A}}
\newcommand{\rlc}{r_{\rm lc}}
\newcommand{\tsd}{t_{\rm em}}
\newcommand{\omegas}{\hat{\omega}_{\rm s}}
\newcommand{\OmegaK}{\Omega_{\rm K}}
\newcommand{\PK}{P_{\rm K}}

\usepackage[T1]{fontenc}
\usepackage{ae,aecompl}

\usepackage{txfonts}

\title[Spin-down of the magnetar in RCW~103]{Ejector and propeller spin-down: How might a superluminous supernova millisecond magnetar become the 6.67~hr pulsar in RCW~103}

\author[W. C. G. Ho and N. Andersson]{Wynn C. G. Ho$^{1,2}$\thanks{Email: \href{mailto:wynnho@slac.stanford.edu}{wynnho@slac.stanford.edu}}, Nils Andersson$^{1}$
\\
$^{1}$Mathematical Sciences and STAG Research Centre, University of Southampton, Southampton, SO17 1BJ, UK \\
$^{2}$Physics and Astronomy and STAG Research Centre, University of Southampton, Southampton, SO17 1BJ, UK \\
}

\date{Accepted 2016 September 9. Received 2016 September 9; in original form 2016 August 10}

\pubyear{2016}

\begin{document}
\label{firstpage}
\pagerange{\pageref{firstpage}--\pageref{lastpage}}
\maketitle

\begin{abstract}
The X-ray source \rcw\ in the supernova remnant \rcwone\ recently exhibited
X-ray activity typical of magnetars, i.e. neutron stars with magnetic fields
$\gtrsim 10^{14}-10^{15}\mbox{ G}$.
However, \rcw\ has an observed period of 6.67~hr, in contrast to magnetars
which have a spin period of seconds.
Here we describe a simple model which can explain the spin evolution of \rcw,
as well as other magnetars, from an initial period of milliseconds that would
be required for dynamo generation of magnetar-strength magnetic fields.
We propose that the key difference between \rcw\ and other magnetars is the
persistence of a remnant disk of small total mass.
This disk caused \rcw\ to undergo ejector and propeller phases in its life,
during which strong torques caused a rapid increase of its spin period.
By matching its observed spin period and $\approx 1-3$~kyr age, we find that \rcw\
has the (slightly) highest magnetic field of all known magnetars,
with $B\sim 5\times 10^{15}\mbox{ G}$, and that its disk had a mass of
$\sim 10^{24}\mbox{ g}$, comparable to that of the asteroid Ceres.
\end{abstract}


\section{Introduction} \label{sec:intro}

The recent detection by \citet{daietal16,reaetal16} of high energy activity
from the neutron star (NS) \rcw\ in supernova remnant \rcwone\ (also known
as SNR G332.4$-$0.4) finally provides insights into its perplexing nature.
\citet{tuohygarmire80} discovered the X-ray point source \rcw\ using
\textit{Einstein},
but no corresponding optical/IR or radio counterpart has been found to date
\citep{tuohyetal83,delucaetal08}, which argues in part against the source
being in a binary system.
The age of \rcwone\ is $\sim 3.3\mbox{ kyr}$ \citep{clarkcaswell76}
or within the range $1.2-3.2\mbox{ kyr}$ \citep{nugentetal84,carteretal97}.
\citet{garmireetal00} find that \rcw\ pulsates with a period of
$\sim 6\mbox{ hr}$, with a more definitive and refined detection of 6.67~hr
determined by \citet{delucaetal06}.
Continued monitoring of \rcw\ yields a constraint on the time derivative of
this period of $\dot{P}\le 1.6\times 10^{-9}\mbox{ s s$^{-1}$}$
\citep{espositoetal11}, which is higher than the $\dot{P}$ of all known
isolated pulsars.
The 6.67~hr period makes \rcw\ a particularly interesting object.

Some characteristics of \rcw\ match those of the central compact object (CCO)
class of NSs, which are found near the center of SNR and
are only seen in X-rays, and thus \rcw\ had been associated with this class
(see \citealt{deluca08,halperngotthelf10,gotthelfetal13}, for review;
see also \citealt{ho13}).
CCOs have an inferred magnetic field $B\sim 10^{10}-10^{11}\mbox{ G}$, and
three CCOs have a measured spin period $P$: two have $P=0.1\mbox{ s}$ and
one has $P=0.4\mbox{ s}$.
If the 6.67~hr period of \rcw\ is ascribed to its spin, then this value is
in sharp contrast to the spin period of CCOs.

However the recent high energy activity is very similar to activity seen in
another class of NSs, that of the magnetars.
Magnetars traditionally include two types of NSs observed at high energies,
anomalous X-ray pulsars (AXPs) and soft gamma-ray repeaters (SGRs), almost
all of which have an inferred magnetic field $B\sim 10^{14}-10^{15}\mbox{ G}$
(see \citealt{mereghettietal15,turollaetal15}, for review).
Thus it is likely that \rcw\ also possesses a magnetic field in this range,
and hereafter we will assume this is the case.
While $B$ may be similar to other magnetars, the 6.67~hr period of \rcw\
(which we will assume is its spin period; see \citealt{daietal16,reaetal16})
is drastically longer than the spin period of other magnetars, which are all
in the range $2-12\mbox{ s}$.
Furthermore, this long period of \rcw\ could be used to argue against the
generation of magnetar-strength magnetic fields via a dynamo mechanism, since
this mechanism requires an initial rapid (possibly millisecond) spin period
\citep{thompsonduncan93,bonannoetal05,spruit09,ferrarioetal15}.

In Section~\ref{sec:model}, we present calculations of a simple scenario
that can describe the spin evolution of the magnetar \rcw, starting from
its birth with a spin period of a millisecond to its current 6.67~hr,
and how \rcw\ is different from other known magnetars.
The scenario is as follows: A supernova gives birth to a rapidly rotating NS.
The rapid spin rate allows us to retain the dynamo mechanism as a viable
means to explain the magnetic fields of \rcw\ and other magnetars
\citep{thompsonduncan93,bonannoetal05,spruit09}.
We also note that our theoretically conceived millisecond magnetar connects
observed magnetars (that are $\gtrsim 10^3\mbox{ yr}$ old) to those thought
to power superluminous supernovae
(see, e.g. \citealt{chatzopoulosetal13,inserraetal13,nicholletal14}).
Next, after an initial epoch during which the immediate environs surrounding
the newborn NS settles into a relatively homogeneous low density plasma and
the magnetic field organizes itself into an ordered dipole field,
we begin at time $t_0$ with a millisecond magnetar
(with $P_0\sim 1\mbox{ ms}$ and $B\sim 10^{14}-10^{15}\mbox{ G}$).
This millisecond magnetar evolves as a standard pulsar, i.e. it loses
rotational energy and slows down as a result of the NS emitting
electromagnetic dipole radiation.
For most known magnetars, this spin-down continues for thousands of years
until their present age and produces NSs that have a spin period of a few
seconds (see after equation~\ref{eq:tmag}), just as observed.
In contrast, we propose that for \rcw, there remained some material that
was not ejected by the supernova
(we estimate a total mass of about that of the asteroid Ceres;
see Sect.~\ref{sec:discuss}), and it forms a remnant disk around the NS
(see, e.g. \citealt{michel88,linetal91,pernaetal14}).
The rapid rotation of the NS causes it to be in an ejector state/phase and
prevents the remnant disk from interacting with the NS
\citep{illarionovsunyaev75}.
The duration of the ejector phase $\tej$ can be hundreds to thousands of
years (see equation~\ref{eq:tej}), and all the while the NS emits dipole
radiation and its spin period increases.
Eventually its rotation becomes slow enough for disk material to couple to
the NS magnetosphere, and the NS transitions to a propeller state/phase.
In this state, matter is expelled by the (still) rapidly rotating NS, and
the resulting spin-down torque on the NS is much stronger than that due to
dipole radiation.  The NS spin period increases at an exponential rate
(see equation~\ref{eq:spinevolprop}) for a short time $\tprop$
(see equation~\ref{eq:tprop}), before reaching spin equilibrium, when
torques on the NS balance.  The result is a slowly spinning, strongly
magnetized NS, like \rcw.

Here we briefly mention previous works which sought to explain the 6.67~hr
period of \rcw\ as its spin period.
\cite{delucaetal06} (see also \citealt{espositoetal11}) ignore the ejector
phase and begin their calculation of propeller phase spin-down at
$P_0=300\mbox{ ms}$,
finding that \rcw\ has $B=5\times 10^{15}\mbox{ G}$ and a remnant disk mass
of $3\times 10^{-5}M_{\sun}$.
\citet{li07} describe an ejector and propeller evolution scenario
and perform Monte Carlo simulations to obtain the magnetar spin period
distribution.
\citet{pizzolatoetal08} consider the torque exerted by a binary companion star
and find that \rcw\ has $B\sim 10^{15}\mbox{ G}$ and is in spin equilibrium.
\citet{ikhsanovetal13} consider \rcw\ to have $B\sim10^{12}\mbox{ G}$
and is accreting from a magnetized remnant disk.
We also note the earlier studies of AXPs and SGRs as normal magnetic field
($\sim10^{12}\mbox{ G}$) NSs that are accreting in the propeller phase, but near
spin equilibrium, from a fossil disk (with constant mass; \citealt{alpar01};
or with decreasing mass; \citealt{chatterjeeetal00,ertanetal09}).
While in the final stages of preparing our work, we became aware of the
work of \citet{tongetal16}, who consider a similar scenario as described
here but obtain a much larger disk mass of $\sim 10^{-5}M_{\sun}$
(see Sec.~\ref{sec:discuss}).

\section{Spin period evolution in ejector and propeller phases}
\label{sec:model}

The scenario for the evolution of the \rcw\ spin period described in
Section~\ref{sec:intro} requires a model for ejector and propeller phases
(defined below).
At early times in the ejector phase, a NS spins down in a similar fashion
to an isolated pulsar, i.e. the NS
emits electromagnetic dipole radiation and loses rotational energy.
This energy loss produces a torque on the NS
\begin{eqnarray}
N_{\rm em} &=& -\frac{2\mu^2\Omegas^3\sin^2\theta}{3c^3}
= -\frac{B^2R^6\Omegas^3\sin^2\theta}{6c^3}=-\beta I\Omegas^3 \nonumber\\
&=& -1.5\times 10^{45}\mbox{ erg } B_{15}^2(P/1\mbox{ ms})^{-3}, \label{eq:nem}
\end{eqnarray}
where $\Omega$ ($=2\pi/P$) is spin frequency,
$\theta$ is the angle between stellar rotation and magnetic axes,
$\beta\equiv2\mu^2/3c^3I=B^2R^6/6c^3I=6.2\times 10^{-12}\mbox{ s }B_{15}^2$,
$B_{15}=B/10^{15}\mbox{ G}$,
and we assume the magnetic dipole moment $\mu=BR^3/2$ and an orthogonal rotator,
i.e. $\sin^2\theta=1$.
We take NS mass, radius, and moment of inertia to be
$M=1.4M_{\sun}$, $R=10\mbox{ km}$, and $I=10^{45}\mbox{ g cm$^2$}$,
respectively.
For simplicity we use the traditional vacuum dipole formula of
\citet{pacini68,gunnostriker69}.  Corrections due to a plasma-filled
magnetosphere and in the $\theta$-dependence only introduce changes of
order unity (see, e.g. \citealt{spitkovsky06,contopoulosetal14}).
Torque on the star is defined by $N=I\dot{\Omegas}$,
and, without additional sources of torque on the pulsar, the resulting
evolution equation for spin frequency is $d\Omegas/dt=-\beta\Omegas^3$,
with solution
\begin{equation}
\Omegas=\Omega_0(1+2\beta\Omegas_0^2t)^{-1/2}=\Omega_0(1+t/\tsd)^{-1/2}
 \quad\mbox{for $t_0<t<\tej$}, \label{eq:spinevol}
\end{equation}
where $\Omegas_0$ ($=2\pi/P_0$) is initial spin frequency and spin-down occurs
on the timescale
\begin{equation}
\tsd = 1/2\beta\Omegas_0^2
 = 2.0\times 10^{3}\mbox{ s }B_{15}^{-2}(P_0/1\mbox{ ms})^{2}. \label{eq:tmag}
\end{equation}
From eq.~(\ref{eq:spinevol}) we see that, in isolation, \rcw\ would spin down
to $P\approx 2\pi(2\beta t)^{1/2}=3.9\mbox{ s }B_{15}(t/1000\mbox{ yr})^{1/2}$,
which coincides with the spin period range $P\approx 2-12\mbox{ s}$ of
other observed magnetars \citep{mereghettietal15,turollaetal15}
but is much shorter than its current spin period of $2.4\times 10^4\mbox{ s}$.
This demonstrates that all magnetars except \rcw\ could simply have spun
down to their current spin period via the torque due to electromagnetic
dipole radiation (equation~\ref{eq:nem}).
For \rcw, dipole radiation torque is too weak, and a stronger, additional or
alternative torque, such as that due to mass accretion, is required to
increase its spin period by its current age of a few thousand years.

Therefore let us suppose that when \rcw\ was first born, it was surrounded by
a disk of material from, e.g. supernova ejecta that did not escape the system
\citep{chevalier89}.
This material cannot interact with the pulsar as long as the pulsar light
cylinder, defined by radius
\begin{equation}
\rlc = c/\Omegas=47.7\mbox{ km }\left(P/\mbox{1 ms}\right),
\end{equation}
is smaller than the magnetosphere, whose radial extent is approximately
\begin{equation}
\rmag = \xi\ralf = \xi\left(\frac{\mu^4}{8GM\dot{M}^2}\right)^{1/7}
 = 7.3\times 10^{5}\mbox{ km }\xi B_{15}^{4/7}\dot{M}_{-12}^{-2/7},
 \label{eq:rmag}
\end{equation}
where $\xi\sim 0.5-1$ (see, e.g. \citealt{ghoshlamb79,wang96}),
the Alfv\'en radius $\ralf$ is derived from balancing ram pressure of
the accreting material with pressure of the pulsar magnetic field
\citep{lambetal73,lipunov92}, $\dot{M}$ is mass accretion rate, and
$\dot{M}_{-12}=\dot{M}/10^{-12}M_\odot\mbox{ yr$^{-1}$}$.
Thus this ejector phase takes place when $\rlc<\rmag$.
The transition between ejector and propeller phases occurs at spin period
\begin{equation}
\Pej = \frac{2\pi}{\Omegaej} = \frac{2\pi\rmag}{c}
 = 15\mbox{ s } \xi B_{15}^{4/7}\dot{M}_{-12}^{-2/7}, \label{eq:pspinej}
\end{equation}
and the duration of the ejector phase $\tej$ can be estimated from
eqs.~(\ref{eq:spinevol}) and (\ref{eq:pspinej}) and is
\begin{equation}
\tej = \tsd\left[\left(\frac{\Omegas_0\rmag}{c}\right)^2-1\right]
 \approx \frac{\rmag^2}{2\beta c^2}
 = 1.5\times 10^{4}\mbox{ yr }\xi^2 B_{15}^{-6/7}\dot{M}_{-12}^{-4/7}.
 \label{eq:tej}
\end{equation}

Once $\rlc>\rmag$, the propeller phase begins, and the total torque on the
star is approximately
\begin{equation}
N = N_{\rm acc}+N_{\rm prop}
 \equiv \dot{M}\rmag^2\OmegaK(\rmag)-\dot{M}\rmag^2\Omegas
 = N_{\rm acc}\left(1-\omegas\right), \label{eq:nprop}
\end{equation}
where $N_{\rm acc}$ is accretion (spin-up) torque and
$N_{\rm prop}$ is propeller (spin-down) torque
(see \citealt{hoetal14}, for derivation;
see also, e.g. \citealt{alpar01,espositoetal11,piroott11};
alternative prescriptions for total torque can be found in,
e.g. \citealt{menouetal99,ertanetal09,parfreyetal16}).
The Kepler orbital frequency $\OmegaK(\rmag)$ at the magnetosphere radius
has the corresponding period
\begin{equation}
\PK(\rmag) = \frac{2\pi}{\OmegaK(\rmag)}
 = \left(\frac{4\pi^2\rmag^{3}}{GM}\right)^{1/2}
 = 9.0\times 10^3\mbox{ s }\xi^{3/2}B_{15}^{6/7}\dot{M}_{-12}^{-3/7}.
 \label{eq:pk}
\end{equation}
The fastness parameter $\omegas$ [$\equiv\Omegas/\OmegaK(\rmag)$;
\citealt{elsnerlamb77}] determines
whether centrifugal force due to stellar rotation ejects matter and spins
down the star (propeller phase with $\omegas>1$) or matter accretes and
spins up the star (accretor phase with $\omegas<1$)
(see, e.g. \citealt{wang95}).
The transition between these two phases ($\omegas\approx 1$) is where
the total torque is approximately zero and the NS is in spin equilibrium
\citep{davidsonostriker73,alparetal82} and occurs at spin period
$P=\PK$.

The evolution equation for spin frequency is obtained by equating
eq.~(\ref{eq:nprop}) to stellar torque $N=I\dot{\Omegas}$, so that
(see also \citealt{alpar01})
\begin{equation}
\frac{d\Omegas}{dt}=-\frac{\dot{M}\rmag^2}{I}\left[\Omegas-\OmegaK(\rmag)\right]
 = -\frac{\Omegas}{\tprop}+\frac{\OmegaK(\rmag)}{\tprop},
\end{equation}
where
\begin{equation}
\tprop\equiv\frac{I}{\dot{M}\rmag^2}
 =96\mbox{ yr }\xi^{-2}B_{15}^{-8/7}\dot{M}_{-12}^{-3/7}.
 \label{eq:tprop}
\end{equation}
We can obtain a simple solution of the evolution equation by assuming
$\mu$ and $\dot{M}$ are constant
(more sophisticated models with $\dot{M}(t)$ can be found in,
e.g. \citealt{chatterjeeetal00,ertanetal09,tongetal16}).
Then the spin frequency as a function of time during the propeller phase,
which starts from the end of the ejector phase at time $\tej$ with spin
frequency $\Omegaej$, is
\begin{equation}
\Omegas=\left[\Omegaej-\OmegaK(\rmag)\right]e^{-(t-\tej)/\tprop}+\OmegaK(\rmag)
 \quad\mbox{for $t>\tej$}. \label{eq:spinevolprop}
\end{equation}

\begin{figure}
 \includegraphics[width=\columnwidth]{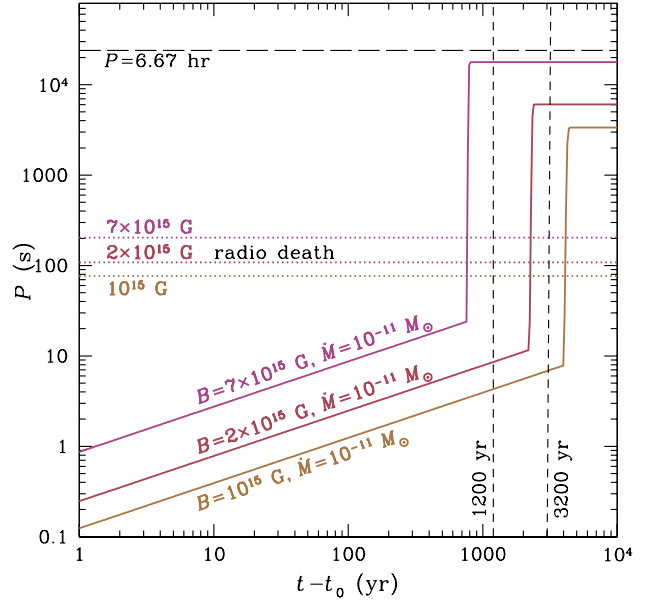}
 \caption{
Spin period as a function of time, starting from ejector phase onset
at $t_0$ with initial period $P_0=1\mbox{ ms}$, for magnetic field $B=1$, 2,
and $7\times 10^{15}\mbox{ G}$ and average mass accretion rate
$\Mdot=10^{-11}M_{\sun}\mbox{ yr$^{-1}$}$.
Horizontal and vertical dashed lines denote the current spin period 6.67~hr
and age range 1200--3200~yr, respectively, of \rcw.
Dotted lines indicate the (theoretically uncertain) death line for pulsar
radio emission with the magnetic fields shown.
}
 \label{fig:spinevol}
\end{figure}

\begin{figure}
 \includegraphics[width=\columnwidth]{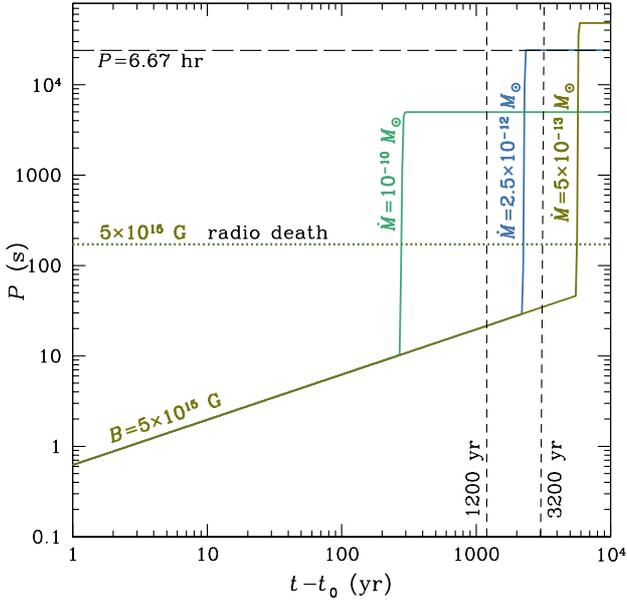}
 \caption{
Spin period as a function of time, starting from ejector phase onset
at $t_0$ with initial period $P_0=1\mbox{ ms}$, for magnetic field
$B=5\times 10^{15}\mbox{ G}$ and average mass accretion rate $\Mdot=10^{-10}$,
$2.5\times 10^{-12}$, and $5\times 10^{-13}M_{\sun}\mbox{ yr$^{-1}$}$.
Horizontal and vertical dashed lines denote the current spin period 6.67~hr
and age range 1200--3200~yr, respectively, of \rcw.
Dotted line indicates the (theoretically uncertain) death line for pulsar
radio emission with $B=5\times 10^{15}\mbox{ G}$.
}
 \label{fig:spinevolmdot}
\end{figure}

Equations~(\ref{eq:spinevol}) and (\ref{eq:spinevolprop}) thus describe the
complete evolution of NS spin frequency (or spin period) through the ejector
and propeller phases, respectively.
Figures~\ref{fig:spinevol} and \ref{fig:spinevolmdot} plot this evolution,
assuming $\xi=1$, an initial spin period $P_0=1\mbox{ ms}$, and different
combinations of magnetic field $B$ and average accretion rate $\Mdot$.
We note that, as long as $P_0\ll\Pej$, the evolution of spin period is
unchanged for any $P_0$, except at very early times.
During the early evolution
(at $t<\tej\sim 10^2-10^3\mbox{ yr}$, depending on $B$ and $\Mdot$;
see equation~\ref{eq:tej}), the NS is in the ejector phase, and
$P\propto t^{1/2}$ (see equation~\ref{eq:spinevol}).
At time $\tej$ when $\rmag=\rlc$, the NS magnetosphere can interact with
the remnant disk, and the NS enters the propeller phase.  The spin period
increases rapidly in this phase
($P\propto e^{t}$; see equation~\ref{eq:spinevolprop}) for a time
$\sim\tprop$ (see equation~\ref{eq:tprop}).
Finally, when $P$ approaches $\PK(\rmag)$ (see equation~\ref{eq:pk}),
propeller and accretion torques balance, so that the total torque on the
star is zero and $P$ is approximately constant, and the NS is in spin
equilibrium.  Figure~\ref{fig:spinevol} shows that, for a given accretion
rate, more strongly magnetized NSs reach longer periods,
while Fig.~\ref{fig:spinevolmdot} shows that, for a given magnetic field,
lower accretion rates produce longer spin period NSs
(see also equation~\ref{eq:pk}).

\begin{figure}
 \includegraphics[width=\columnwidth]{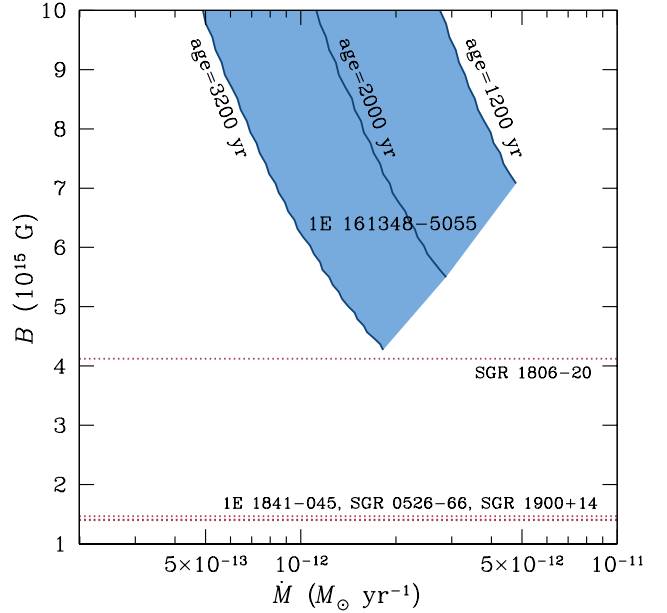}
 \caption{
Constraints on magnetic field $B$ and average mass accretion rate $\Mdot$ of
\rcw.  The shaded region denotes $B$ and $\Mdot$ values which produce a NS
with a spin period of 6.67~hr after 1200--3200~yr from birth.
Horizontal dotted lines indicate the magnetic field of particular magnetars
inferred from their spin period $P$ and spin period time derivative $\dot{P}$
(values taken from the ATNF Pulsar Catalogue;
\citealt{manchesteretal05}) and $B=6.4\times 10^{19}\mbox{ G }(P\dot{P})^{1/2}$.
}
 \label{fig:rcwbm}
\end{figure}

Since we know the spin period $P$ and approximate age of \rcw,
only particular combinations of magnetic field and average mass accretion
rate will satisfy eq.~(\ref{eq:spinevolprop}), i.e. for
\begin{equation}
\ln\frac{\Omegaej}{\Omegas-\OmegaK}=\frac{\left|\mbox{age}-\tej\right|}{\tprop},
\end{equation}
where we take $\Omegaej-\OmegaK\approx\Omegaej$,
the left-hand side must equal the right-hand side
and spin frequency and age are set by $\Omegas=2\pi/(6.67\mbox{ hr})$ and
age =~1200--3200~yr, respectively.
The values of $B$ and $\Mdot$ which satisfy the above are indicated by
the shaded region in Fig.~\ref{fig:rcwbm},
along with the magnetic field of several magnetars, inferred from their
$P$ and $\dot{P}$ (values taken from the ATNF Pulsar
Catalogue\footnote{http://www.atnf.csiro.au/research/pulsar/psrcat/};
\citealt{manchesteretal05};
see also McGill Online Magnetar
Catalog\footnote{http://www.physics.mcgill.ca/~pulsar/magnetar/main.html};
\citealt{olausenkaspi14}), the highest
of which is $4\times 10^{15}\mbox{ G}$ for SGR~1806$-$20.
If \rcw\ has a slightly higher field of $\approx 5\times 10^{15}\mbox{ G}$
than SGR~1806$-$20 and is $\approx 2300\mbox{ yr}$ old,
then it only requires an average accretion rate of
$\approx 2.5\times 10^{-12}M_{\sun}\mbox{ yr$^{-1}$}$ to spin it down
to a period of 6.67~hr (see also Fig.~\ref{fig:spinevolmdot}).
If the accretion rate is much lower or higher (at this $B$), then \rcw\
would be in the ejector phase or in spin equilibrium, respectively,
with a spin period much shorter than 6.67~hr in both cases.

\section{Discussion} \label{sec:discuss}

Recent observations by \citet{daietal16,reaetal16} of the X-ray source \rcw\
in SNR \rcwone\ strongly suggest it is a magnetar
(NS with $B\gtrsim 10^{14}\mbox{ G}$) with an extremely long spin period
$P=6.67\mbox{ hr}$, in contrast to all other known magnetars which have
2--12~s spin periods.
The long spin period of \rcw\ might argue against dynamo generation of
magnetic fields because of a requirement for fast (millisecond) initial
spin periods,
and there is insufficient time for \rcw\ to lose enough rotational energy
via conventional electromagnetic dipole radiation.

Here we demonstrate, using a simple model with simple assumptions,
that the spin period of \rcw\ can increase from
milliseconds to 6.67~hr over its 1.2--3.2~kyr lifetime by evolving through
ejector and propeller phases while undergoing accretion from a disk.
Our calculations show that a young NS, such as \rcw, can spend quite a long
time $\tej$ in the ejector phase, and thus this phase should not be neglected.
The requisite disk might have remained bound to the NS during its formation
in a (superluminous) supernova and may have significant impact even when it
has very low total mass, which we estimate to be
$\Delta M\sim 10^{-12}M_{\sun}\mbox{ yr$^{-1}$}\times 10^3\mbox{ yr}
=10^{-9}M_{\sun}$.
This disk may still be present or has been completely accreted/dissipated.
In the case of the former, since \rcw\ has evolved to be near spin equilibrium
(when the total torque on the star is approximately zero), the long-term
$\dot{P}$ is very low and could satisfy the observed constraint of
$\dot{P}<10^{-9}\mbox{ s s$^{-1}$}$ obtained by \citet{espositoetal11}.
\rcw\ may on occasion accrete more or less material, which could explain the
variability seen in X-rays \citep{gotthelfetal99,delucaetal06}.
In fact, our derived $\dot{M}\sim 10^{-12}M_{\sun}\mbox{ yr$^{-1}$}$
corresponds to a luminosity $L=GM\dot{M}/R\sim 10^{34}\mbox{ erg s$^{-1}$}$,
which is on the order of that observed \citep{delucaetal06}.

If the disk is no longer present (with the observed X-ray variability due
to typical magnetar variability), then the dipole radiation torque yields
$\dot{P}\sim 10^{-13}-10^{-12}\mbox{ s s$^{-1}$}$,
well below the observed constraint.
We also note that, while the exact nature of the mechanism that causes radio
emission is uncertain, it is thought that there exists a ``death line''
which demarcates when observable radio emission ceases
\citep{rudermansutherland75,bhattacharyaetal92}.
This death line is shown by dotted lines in Figs~\ref{fig:spinevol} and
\ref{fig:spinevolmdot}.
It is clear that, while the NS is in the ejector phase, its spin
period is below the death line, and as such, it could emit as a radio pulsar.
However after transition to propeller phase, the spin period rapidly
increases above the death line.  Thus once the accretion disk material is
exhausted and the propeller phase ceases, the NS will not emit as
a radio pulsar.

Finally, our results suggest a possible unified formation scenario for
various classes of observed NSs.
This scenario is schematically described in Table~\ref{tab:nsform}
and depends on total accreted mass and time spent accreting
following a chaotic and turbulent supernova
(a scenario that includes more classes but is a function of accretion rate
from a fallback disk is proposed in \citealt{alpar01}).
For short duration accretion (hours to possibly days) of a large amount
of mass ($\gtrsim 10^{-4}M_{\sun}$), accreting matter can build up on the
NS surface so fast that the magnetic field is buried temporarily.
Once accretion slows or stops, the magnetic field re-emerges on a timescale
of $\sim 10^2-10^4\mbox{ yr}$, depending on burial depth, and this
increasing surface field could explain properties of CCOs \citep{ho11,ho13}.
For small to no accretion, we transition from CCO formation to pulsars that
have possible signatures of magnetic field growth, e.g. their braking index,
to the majority of isolated radio pulsars \citep{ponsetal12,ho15}.
In the case of accretion for long durations, the magnetic field will not
be buried if the total mass is small.
For large total mass
(e.g. $10^{-5}M_{\sun}$, like that of the disk seen around magnetar
4U~0142+61; \citealt{wangetal06}),
NSs with $B\sim 10^{13}\mbox{ G}$ could end up with $P\sim 10\mbox{ s}$ in
$\lesssim 10^5\mbox{ yr}$.
However, \citet{pernaetal14} show that it is extremely difficult to retain
such amounts for long durations during a supernova.
For \rcw, only a small amount of mass ($\Delta M\sim 10^{-9}M_{\sun}$)
needs to be retained following its supernova.
Thus \rcw\ is possibly a very special system, and the interaction of its
magnetic field with this small mass over a thousand years is what leads
to its long spin period of 6.67~hr.

\begin{table}
\caption{Cases for formation of different neutron star populations}
\label{tab:nsform}
\begin{tabular}{ccc}
\hline
 & \multicolumn{2}{c}{time spent accreting} \\
 & short & long \\
small $\Delta M$ & radio pulsar & \rcw \\
large $\Delta M$ & CCO & improbable formation \\
\hline
\end{tabular}
\end{table}

\section*{Acknowledgements}
The authors thank the anonymous referee for helpful comments.
WCGH and NA acknowledge support from the Science and Technology Facilities
Council (STFC) in the United Kingdom.

\label{lastpage}
\end{document}